\documentclass[a4paper]{PoS}
\usepackage{url}

\title{Calibration of the Cherenkov Telescope Array}

\ShortTitle{Calibration of CTA}

\author{Markus Gaug$^{a}$, \speaker{Michael Daniel}$^{b}$, David Berge$^{c}$, Raquel de los Reyes$^{d}$, Michele Doro$^{e}$, Andreas Foerster$^{d}$, Maria~Concetta Maccarone$^{f}$, Dan Parsons$^{d}$,
        Christopher van Eldik$^{g}$, for the CTA Consortium\footnote{Full consortium author list at http://cta-observatory.org}\\
        E-mail: \email{markus.gaug@uab.cat, michael.daniel@liverpool.ac.uk}

{\footnotesize
$^{a}$            Unitat de F\'isica de les Radiacions, Departament de F\'isica, and CERES-IEEC, Universitat Aut\`onoma de Barcelona, E-08193 Bellaterra, Spain.\\
$^{b}$            Department of Physics, University of Liverpool, Liverpool, L69 7ZE. UK.\\
$^{c}$            GRAPPA, Anton Pannekoek Institute for Astronomy, University of Amsterdam,  Science Park 904, 1098 XH Amsterdam, The Netherlands.\\
$^{d}$            Max-Planck-Institut f\"ur Kernphysik, P.O. Box 103980, D-69029 Heidelberg, Germany.\\
$^{e}$            University and INFN Padova, Via Marzolo 8, 35131 Padova, Italy. \\
$^{f}$            INAF -- IASF Palermo, Via U. La Malfa 153, I-90146 Palermo, Italy.\\
$^{g}$            Erlangen Centre for Astroparticle Physics, Universit\"at Erlangen N\"urnberg, Germany. }

}

\abstract{The construction of the Cherenkov Telescope Array, consisting of two observatories designed to observe the very high energy gamma-ray sky with unprecedented sensitivity and precision, is expected to start soon. We will present the baseline methods and their extensions currently foreseen to calibrate the observatory. These are bound to achieve the strong requirements on allowed systematic uncertainties for the reconstructed gamma-ray energy and flux scales, as well as on the pointing resolution, and on the overall duty cycle of the observatory.
Onsite calibration activities are designed to include a robust and efficient calibration of the telescope cameras, and various methods and instruments to achieve calibration of the overall optical throughput of each telescope, leading to both inter-telescope calibration and an absolute calibration of the entire observatory. One important aspect of the onsite calibration is a correct understanding of the atmosphere above the telescopes, which constitutes the calorimeter of this detection technique. It is planned to be constantly monitored with state-of-the-art instruments to obtain a full molecular and aerosol profile up to the stratosphere. In order to guarantee the best use of the observation time, in terms of usable data, an intelligent scheduling system is required, which gives preference to those sources  and observation programs that can cope with the given atmospheric conditions, especially if the sky is partially covered by clouds, or slightly contaminated by dust. Ceilometers in combination with all-sky-cameras are plannned to provide the observatory with a fast, online and full-sky knowledge of the expected conditions for each pointing direction. For a precise characterization of the adopted observing direction, wide-field optical telescopes and Raman Lidars are planned to provide information about the height-resolved and wavelength-dependent atmospheric extinction, throughout the field-of-view of the cameras. }

\FullConference{The 34th International Cosmic Ray Conference,\\
		30 July- 6 August, 2015\\
		The Hague, The Netherlands}

\begin{document}

\newcommand\T{\rule{0pt}{2.6ex}}        
\newcommand\TT{\rule{0pt}{2.1ex}}        
\newcommand\B{\rule[-1.5ex]{0pt}{0pt}}  
\newcommand\BB{\rule[-1.0ex]{0pt}{0pt}} 
\section{Introduction}

The observation of very high energy gamma-rays using Cherenkov light produced in extended air showers has become a standard tool of 
modern astronomy over the last decade~\cite{hillas2013}. Based on the success of the present generation of ground-based Cherenkov telescope arrays
(H.E.S.S., MAGIC and VERITAS\,\footnote{%
Additional information on those experiments can be found at 
\url{www.mpi-hd.mpg.de/hfm/HESS/}, \url{wwwmagic. mppmu.mpg.de} and \url{veritas.sao.arizona.edu}, 
respectively.}), the future Cherenkov Telescope Array 
 (CTA)~\cite{ctaconcept} 
 will provide the astrophysical community with one mature open-access gamma-ray observatory in both hemispheres 
for the observation of gamma-rays with energies from a fews tens of GeV to beyond 100~TeV
with unprecedented sensitivity and angular and energy resolution~\cite{hinton2013,bernloehr2013}. 
To realise these goals, 3 sizes of telescopes are planned: 23~m diameter Large-Sized-Telescopes (LSTs), designed to provide the CTA with a low energy threshold, 
12~m diameter Medium-Sized-Telescopes (MST) which should provide the majority of the improvement in flux sensitivity in the 100~GeV to 1~TeV energy range, and 
a large number of Small-Sized-Telescopes (SSTs) to extend the high-energy reach of the CTA. 

The current generation of Imaging Atmospheric Cherenkov Telescopes (IACTs) has been designed with the main
aim of discovering new types of Very High Energy (VHE) gamma-ray emitters in the universe. 
Meanwhile, current instruments are starting to be limited by systematic uncertainties: the necessary calibration precision for the used \textit{calorimeter} (the atmosphere) 
and the Cherenkov \textit{light detection instruments} (the telescopes) had not always been included in their design.
What is more, given the continuing success of these IACTs, facilitated by frequent improvements of sensitivity, 
an exhaustive optimization of the full analysis duty cycle has not been a major issue until recently. 
There is still a potential to recover around 20--30\% of the effective duty cycle, now lost either at the moment when strong data selection 
is required to guarantee data samples with stable systematic uncertainties for a complete spectral and morphological analysis, 
or already during observation, when the telescopes stop data taking under sub-optimal observing conditions\footnote{%
Under these circumstances, \textit{optimal atmospheric conditions} 
are practically identical with \textit{photometric nights}. 
}. 
Some of these instruments 
have learnt to continuously monitor the optical throughput of the telescopes~\cite{bolzphd,kellermann2012} as well as the properties of the atmosphere above them~\cite{fruck2013}, 
and to correct the pointing resolution offline to arcsecond precision scales~\cite{gillessenicrc}. 

Along with the discovery and subsequent establishment of VHE gamma-ray source populations, new questions have arisen, involving the interpretation of source spectra, 
the detection of flux variations and morphology studies of extended sources. Several recent discoveries by H.E.S.S., MAGIC and VERITAS are hence limited by the 
systematic uncertainties with which the instrument response is known. What was good enough at the beginning for the discovery of sources, is now starting to limit performance (see e.g.~\cite{magicsegue,magiccrabnebula,hessebl}). 

The CTA, in turn, is expected to dedicate a significant fraction of its lifetime to \textit{population studies} and \textit{precision measurements}. It will typically resolve spectral 
features such as the location of Inverse Compton peaks, spectral breaks and cutoffs (mainly in the tens of GeV regime and the tens of TeV regime), the imprint of the 
extinction of gamma-rays by the interaction with extragalactic background light on the received energy spectra, and possibly even spectral lines. 
It will also resolve the morphology of sources to unprecedented precision, and determine the precise location of VHE gamma-ray emission.

The physics cases for CTA have been studied 
and boiled down to a series of \textit{high-level requirements}, of which several concern the precision with 
which the physical properties of the incident gamma-rays must be known: 

\begin{itemize}
\item \vspace{-2.5mm}Energy scale: The systematic error in the overall energy scale must be below 10\%.
\item \vspace{-2.5mm}Source localization: The precision with which point sources can be located must be better than 5$^{\prime\prime}$ per axis with the goal of 3$^{\prime\prime}$ per axis.
\item \vspace{-2.5mm}Availability: 100\% of all feasible operational time must be available for observation.
\end{itemize}
\vspace{-2.5mm}
Further requirements have been established as \textit{performance requirements}, 
 the following of which especially affect the continuous online calibration efforts: 

\begin{itemize}
\item \vspace{-2.5mm}Cherenkov light intensity: The systematic error on the measurement of the absolute intensity of the Cherenkov light (post-calibration) at the position of each telescope must be $<$8\%, with the goal of 5\%.
\item \vspace{-2.5mm}Effective areas: The uncertainty on the effective area, well above threshold, must be $<$12\%, with a goal of 8\%.
\item \vspace{-2.5mm}Exposure:   The integrated exposure (well above threshold), on a given target must be known to better than 15\%. 
\end{itemize}
\vspace{-2.5mm}
Some of these requirements can be fulfilled using methods which are state-of-the-art in some of the current IACTs, while others require innovation beyond the current state-of-the-art. 
Especially the maximum availability, and the requirement on the systematic error of the energy scale are quite challenging.

\section{General Strategy}

Experience with the current generation of IACTs (H.E.S.S., MAGIC and VERITAS) has shown that the following \textit{baseline of calibration methods} 
can achieve about 15\% systematic uncertainty for the absolute energy scale~\cite{magicperformance,magicperformance2} and 
10\%--20\% -- depending on the energy range -- for the reconstructed flux~\cite{magicperformance,magicperformance2,hessperformance} (an additional systematic uncertainty of 
$\pm$(0.1--0.15) on the slope of reconstructed power-law spectra is assumed), both 
for quality-selected data, i.e.\ after removal of data taken under non-optimal atmospheric conditions:
\begin{itemize}
\item \vspace{-2.5mm}Analysis of regularly taken single photo-electron spectra~\cite{hesscalibration} or calibration pulses interlaced with air shower 
data taking in combination with the photon statistics methods~\cite{mirzoyan1997} for the camera pixel calibration~\cite{hanna2010},
\item \vspace{-2.5mm}Analysis of muon rings and selected cosmic ray images to calibrate the optical throughput of the individual telescopes~\cite{bolzphd,hofmann2003},
\item \vspace{-2.5mm}Selection of acceptable atmospheric conditions with parameters based on the trigger rates~\cite{hahn2014} in order to control systematic errors on the energy scale to an acceptable level,
\item \vspace{-2.5mm}The recent introduction of standard energy and effective area correction by using a continuously run LIDAR~\cite{fruck2013}.
\end{itemize} 
\vspace{-2mm}
Most of the residual systematics of the current generation instruments are due to un-monitored and un-corrected changes in atmospheric 
conditions, but also un-simulated long-term degradation of mirrors, cameras and of the telescope structure play a role. 
The baseline methods alone do not yet guarantee that the requirements for the CTA are always met, and moreover come along with an unacceptably high data loss rate. On top of that, 
 current IACTs are sensitive only for a part of the energy spectrum covered by the CTA, whereas calibration becomes more and more difficult towards the very lower and 
the very upper end of the energy range, either due to higher and higher atmospheric shower heights, or due to less and less event statistics. 
Successful calibration of the CTA requires hence methods and instruments \textit{outperforming the baseline established by the current IACTs}~\cite{gaugSPIE2014}. 
For implications on and error budgets of the camera calibration, a separate proceeding is available at this conference~\cite{daniel2015}.

On the contrary to the above said, the systematic uncertainty for the localization of a point source of the order of several arc-seconds has already been achieved in current IACTs
by the H.E.S.S. collaboration~\cite{gillessenicrc}, at least within an array of equal telescope sizes. A more detailed discussion on achievable pointing precision of the CTA are presented elsewhere 
in this conference~\cite{eschbach}. 

\section{Calibration of the Optical Throughput}

A collaboration-wide effort was made to establish the feasibility of an optical throughput calibration scheme based on muon rings (see e.g.~\cite{brown,toscano}).
These studies led to the insight that a muon calibration scheme seems viable for all telescopes, using regular data taken close to contemporary with normal science observations, 
but improving the currently applied technique.
Slightly adapted trigger thresholds (possibly adjusted to the expected shapes of muon images) can be necessary for the smallest telescopes at least, as well as sufficiently efficient flagging 
of muon rings which have triggered only one telescope, in order not to get lost by the stereo coincidence trigger. 
Additionally, the telescope and camera components must be designed such that the transmitted part of the muon spectrum below 290~nm becomes negligible, in order to ensure 
sufficient match of the received Cherenkov light spectra from local muons and remote gamma-ray showers.

The precision of muon calibration can then be safely estimated to about 2--3\% systematic uncertainties for any achromatic degradation of the optical throughput 
for Cherenkov photons in the wavelength range between 290~nm and 700~nm. 
In the case of (expected long-term) wavelength-dependent degradation of the optical elements of a telescope, 
the correction applied from the measured efficiency to muon rings might result 
in an over-corrected efficiency to Cherenkov light from gamma-ray showers. 
This over-correction may amount to $\lesssim$13\%. 
It is hence essential to determine the chromaticity of any degradation of optical elements from time to time, e.g. once a year. 

An option for the wavelength-dependent calibration of each telescope can be the use of a calibrated light source flashing the telescopes from a distance of $\gtrsim$100~m, 
the so-called \textit{Illuminator}.
Also light flashers mounted on Unmanned Air Vehicles (UAVs) flashing the telescopes from above~\cite{matthews2013}, 
and calibrated lasers~\cite{gaugjinst} are an option.

Cross calibration of telescope response efficiencies through the use of cosmic ray images has been shown to be a robust approach 
enabling calibration independent  of many different hardware technologies. 
Relative calibration through pairwise comparisons ensures that multiple independent measurements over-determine the system of unknown parameters, 
leading to an overall precision at the 1--3\% level after reasonable data collection times~\cite{mitchell2015}.

\section{Atmospheric Calibration}

Once the telescopes' response to light is determined to better than 4--5\% systematic uncertainty, the impact of the atmosphere must be understood to a level of 8--9\%,
in order to still meet the required 10\% systematic uncertainty on the energy scale of the CTA. 

Simulation codes such as CORSIKA and KASKADE-C++ agree within $\sim$5\% in their prediction for the absolute light yield at ground level and its radial distribution, 
excluding atmospheric effects~\cite{bernloehr2013}. For CTA in turn, it is desirable to understand the predicted light yield in MC simulations to about 2\%. In addition to that, 
desired simplifications in the Monte-Carlo simulation of the air showers lead to additional uncertainties of the order of $\lesssim$2\%. 

At mid-latitudes, seasonal variation of the Cherenkov light yield can be as large as 25\%, mostly due to the difference in the height of shower maximum coupled with the height-dependent threshold for Cherenkov light emission~\cite{bernloehr2000}. 
If measurements or predictions of the atmospheric profile with a precision of about 1~g/cm$^2$ at a height resolution of 20~m are available, about 2.5\% 
uncertainty on the relative radiation length of electrons and gamma rays, and hence the relative uncertainty on the shower energy, which is of the same order of magnitude, can be assumed. 
Since the Cherenkov angle also depends on the molecular density profile, the central Cherenkov light density $\rho_c$ for vertical showers follow~\cite{bernloehr2000}:
\begin{equation}
\rho_c \propto (h^*_\mathrm{med} - h_\mathrm{obs})^{-2} 
\end{equation}
where $h^*_\mathrm{med}$ is the median height of emission for Cherenkov light near the core, and $h_\mathrm{obs}$ the observatory altitude. 

Studies carried out for MAGIC have shown that an rms of about 2~g/cm$^2$ between assumed and true density profile
leads to differences in $\rho_c$ of about 0.5\%, 1.5\%, 3.5\% and 6\% for gamma-ray energies of 20~GeV, 200~GeV, 3~TeV and 70~TeV, respectively. 
This geometrical effect adds to an additional, but smaller, error due to the mis-reconstructed molecular extinction of Cherenkov light.

In order to reduce the errors from the insufficient understanding of the molecular component of the atmosphere to acceptable sizes of $\sim$1~g/cm$^2$ 
it is therefore important to start a dedicated radiosonde campaign, once the CTA sites are selected and gather data to validate global data assimilation systems, like the GDAS~\cite{PAO2012-2}.

Modern radiosondes can also measure the concentration of ozone and will allow the uncertainty due to absorption of light by ozone to be limited to less than 1\%, 
if an ozone climatology is later established. Molecular extinction of Cherenkov light should be controlled to the same precision, given the aforementioned precision 
in mass overburden and height. 

The main contribution to systematic uncertainties stems from the contribution of aerosols and clouds and can show variability on time scales of tens of minutes. Consequently, aerosol extinction needs to be measured/monitored on these time scales, with a precision of $\lesssim$2\%, 
and a height resolution of the order of one radiation length, i.e. $\sim$40~g/cm$^2$. Only the fine-structure of the nocturnal boundary layer does not need to be resolved.

Several CTA groups have started with the design of Raman LIDARs~\cite{dorolidars,arcadelidar} early on in the project, in order to continuously measure and monitor the extinction of Cherenkov light due to aerosols to a precision better than 2\%. LIDAR measurements are desirable since they can identify both ground-level aerosols, where extinction corrections of air shower data are straight-forward~\cite{garrido2013}, and the vertical structure of higher altitude aerosols (e.g. due to distant biomass burning, desert dust intrusion, cirrus or volcanic eruptions) where shower image profiles are significantly distorted and where a correction is more difficult, but nevertheless possible~\cite{fruck2013}. Stratospheric aerosols in turn, may cause a significant obscuration of the star light after strato-volcanic eruptions, but only influence the Cherenkov light from air showers of exceptionally large heights of shower production~\cite{bernloehr2000} and hence only at the energy threshold of the CTA. Hence, relying only on photometry of reference stars may introduce a bias which needs to be corrected for. It is nevertheless useful for calibration and monitoring, 
e.g. with a small robotic optical telescope equipped with a filter wheel, as UVscope~\cite{maccarone2011}, a sophisticated, portable small-aperture multi-pixels photon detector. 

The FRAM telescope~\cite{prouza2010}, used at the Pierre Auger Observatory, is able to perform this procedure over a large field-of-view, of the order of 10$^\circ \times $10$^\circ$, and fine resolution (0.03$^m$), which is very useful for determining the extinction across the field-of-view. Studies have started to determine the impact of cirrus clouds covering parts of the FOV of a Cherenkov camera on the angular and energy resolution, and how to correct for the respective biases offline, once a FRAM picture is available and the altitude of the cloud is determined, e.g. by the Raman LIDARs.

Scattered Cherenkov light plays only a minor role for IACTs~\cite{bernloehr2000}. 
A value of $<$1\% additional contribution of light is expected for the large-size telescopes, and $<$2\% for the small-size telescopes, which will have a 10$^\circ$ field-of-view.

\section{Intelligent target selection}

In order to enhance the effective duty cycle of the CTA, an intelligent scheduling system is planned, which allows to prefer observation of sources visible under good atmospheric conditions over those covered by clouds or aerosol layers. This is especially important in the case of cirrus clouds, which rarely cover the entire sky. Given that the height of an aerosol layer determines the energy threshold of the system~\cite{garrido2013}, and cirrus will impact its angular resolution, such an intelligent scheduling system shall be able to judge at each moment whether the observation requirements are still met, or whether a different source with less strict requirements should be observed instead. Only when optimal observation conditions are not possible, should atmospheric corrections be applied.

For such pointing forecasts, scanning instruments, and/or All-Sky-Cameras~\cite{fruckjinst} are suitable. 
Active scanning instruments should employ wavelengths that do not interfere with the CTA cameras, such as commercial ceilometers (operating at 905~nm or 1064~nm). 
All-Sky Cameras, which have been largely used during the CTA site selection process~\cite{fruckjinst}, 
are becoming more and more the standard tools for cloud detection at world-class astronomical observatories. These highly sensitive devices are able to detect even fine cirrus, by comparing detected light fluxes from stars with their catalog values. Exposure times of less than 1~minute are possible, and provide almost contemporaneous cloud maps. 
The cloud height cannot be measured accurately however, a task left to the ceilometers.



\section*{Acknowledgements}

We gratefully acknowledge support from the agencies and organizations 
listed under Funding Agencies at this website: \url{http://www.cta-observatory.org/}.

\end{document}